\documentclass[aps,prl,twocolumn,groupedaddress,showpacs]{revtex4}

\usepackage{graphicx}
\begin{document}
\title{Quantum tunneling in a three dimensional network of exchange
coupled single-molecule magnets}

\author{R. Tiron$^1$, W. Wernsdorfer$^1$, N. Aliaga-Alcalde$^2$, D.N.
Hendrickson$^3$, G. Christou$^2$}

%\thanks{}
\affiliation{
$^1$Lab. L. N\'eel, associ\'e \`a l'UJF, CNRS, BP 166,
38042 Grenoble Cedex 9, France\\
$^2$Dept. of Chemistry, Univ. of Florida,
Gainesville, Florida 32611-7200, USA\\
$^3$Dept. of Chemistry and Biochemistry, Univ. of California at San
Diego, La Jolla, California 92093-0358, USA
}

\date{\today}

\begin{abstract}
A Mn$_4$ single-molecule magnet (SMM) is used to show that quantum tunneling
of magnetization (QTM) is not suppressed by moderate three dimensional
exchange
coupling between molecules. Instead, it leads to an exchange bias of the
quantum resonances which allows precise measurements of the effective exchange
coupling that is mainly due to weak intermolecular hydrogen bounds. 
The magnetization
versus applied field was recorded on single crystals of [Mn$_4$]$_2$ using an
array of micro-SQUIDs. The step fine structure was
studied via minor hysteresis loops.
\end{abstract}


\maketitle

Single-molecule magnets (SMM), such as 
Mn$_{12}$, Mn$_{4}$ and Fe$_8$
~\cite{Christou00,Sessoli93b,Sessoli93,Aubin96,Boskovic02}, 
have become model systems to study quantum tunneling of magnetization
(QTM)~\cite{Friedman96,Thomas96,Sangregorio97,Aubin98,Aubin98b,WW_Science99}.
These molecules comprise several
magnetic ions, with their spins coupled by strong exchange interactions
to give a large effective spin. The molecules are regularly
assembled in large crystals where often all the molecules have the same
orientation. Hence, macroscopic measurements can give direct access
to single molecule properties. Many non-magnetic atoms surround the
magnetic core of each molecule. Exchange interactions between 
molecules are therefore relatively
weak and have been neglected in most studies.

Recently, the study of a dimerized SMM [Mn$_4$]$_2$ showed
that intermolecular exchange interactions are not 
negligible~\cite{WW_Nature02}.
This compound belongs to the [Mn$_4$O$_3$Cl$_4$(O$_2$C{\bf R})$_3$(py)$_3$]$_2$
family,
with {\bf R} = CH$_2$CH$_3$ and it will be named in the following
as compound {\bf 1}. The spins of the two Mn$_4$ molecules are
coupled antiferromagentically. Each molecule acts as a bias on its neighbor,
the quantum tunneling resonances thus being shifted with 
respect to the isolated
SMM. In this letter we show that even in three-dimensional networks of
exchange coupled SMMs, ordering effects do not quench tunneling.

We selected a dimerized SMM [Mn$_4$]$_2$,
called compound {\bf 2}. The molecule belongs to the same
family as compound {\bf 1}, however
{\bf R} = CH$_3$. Because this substituent has a
smaller volume than {\bf R} = CH$_2$CH$_3$, molecules are packed
closer together. This leads to stronger interdimer
interactions, which are negligible in compound {\bf 1}.
The preparation, X-ray structure and detailed physical characterization
have been reported elsewhere~\cite{Hendrickson92,Aliaga_Alcade}. 
Compounds {\bf 1} and {\bf 2} crystallize in the
hexagonal space group $R3$(bar) with two Mn$_4$
molecules per unit cell lying head-to-head on a crystallographic S$_6$
symmetry axis (Fig. 1). Each monomer Mn$_4$ has a ground state spin
S = 9/2. The Mn-Mn distances and the Mn-O-Mn
angles are similar and the uniaxial anisotropy constant is expected to be
the same for the two dimer systems. These dimers are held together via six
C$-$H$\cdot\cdot\cdot$Cl hydrogen bonds between the pyridine (py) rings
on one molecule
and the Cl ions on the other and one Cl$\cdot\cdot\cdot$Cl Van der Waals interaction
(Fig. 1a). These interactions lead to an antiferromagnetic
superexchange interaction between the two Mn$_4$ units of a 
dimer~\cite{WW_Nature02}.

\begin{figure}
\includegraphics[width=.46\textwidth]{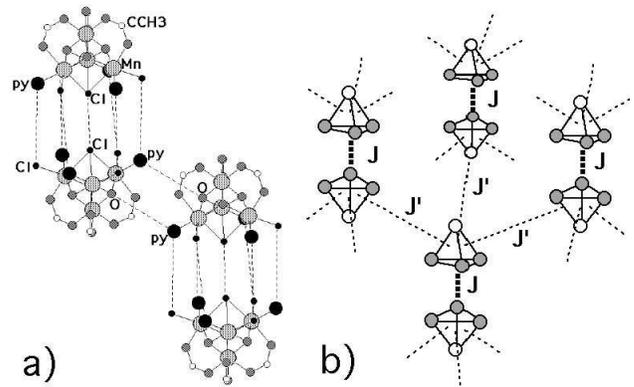}
\caption{(a) X-ray crystal structure of the Mn$_4$ molecule;
the two molecules of a dimer are held together by six
hydrogen bonds (3.6 \AA) between the pyridine rings (py) and the Cl
ions, and one Cl$\cdot\cdot\cdot$Cl van der Waals bond (3.74 \AA). Two neighboring 
dimers interact
via two hydrogen bonds (3.23 \AA) between the py and the O ion.
(b) Schematic view of the exchange coupled 
network of Mn$_4$ molecules. Each Mn$_4$ molecule
(schematized by the Mn$_4$ tetrahedron), is exchange 
coupled to the Mn$_4$ of the dimer ($J$) and to three molecules 
of nearby dimers ($J'$).}
\label{fig1}
\end{figure}

Owing to the S$_6$ symmetry of [Mn$_4$]$_2$,
each Mn$_4$ is close to three neighboring Mn$_4$ molecules of the three
neighboring [Mn$_4$]$_2$ (Fig. 1b). There are hydrogen bonds between the
pyridine (py) rings of the molecules  and the O ions of the other three
neighboring molecules. The C$-$H$\cdot\cdot\cdot$O
hydrogen bonds between [Mn$_4$]$_2$ dimers have C$\cdot\cdot\cdot$O
distances and C$-$H$\cdot\cdot\cdot$O angles of 4.05 \AA\ and
124.85$^{\circ}$,
respectively. The interactions between the dimers are expected to be
antiferromagnetic and weaker than the interdimer interactions.
The two different antiferromagnetic couplings, the stronger one inside
the dimer and the weaker one between the dimers, make this system an
interesting candidate for studying the QTM in a three
dimensional network of exchange coupled SMMs.

\begin{figure}
\begin{center}
\includegraphics[width=.42\textwidth]{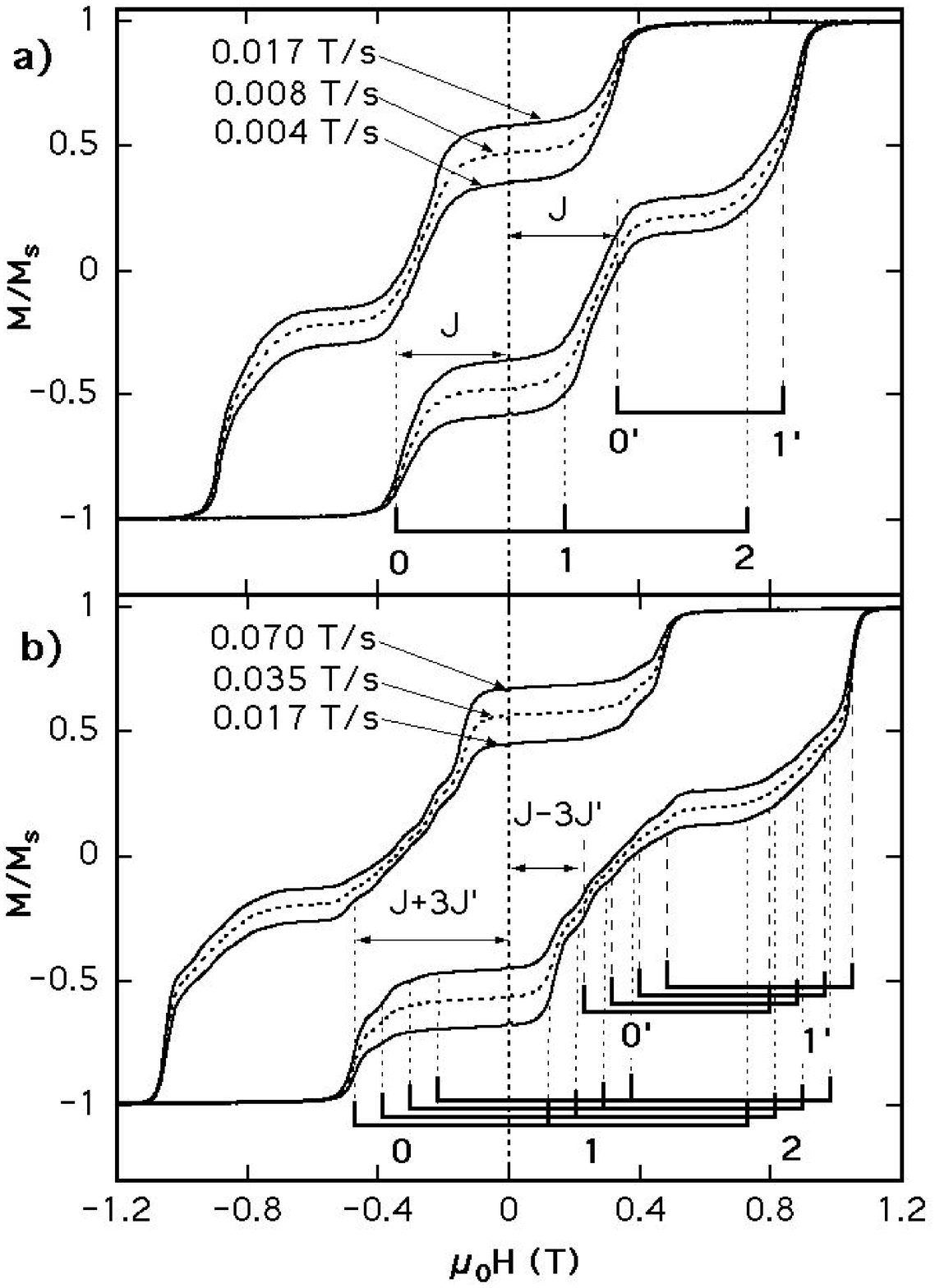}
\caption{ Hysteresis loops for compounds {\bf 1}
(a) and {\bf 2} (b), measured at different sweep rates. If 
the spin of
one molecule is in the -9/2 state, the resonance
positions of the other molecule 
are shifted towards negative fields.
The comb (0,1,2) represents the resonances of one molecule
from -9/2 to +9/2 (0), from -9/2 to +7/2 (1), and 
from -9/2 to +5/2 (2). 
If the spin of the other molecule
is in the +9/2 states, the resonances are shifted 
towards positive fields, indicated by the
comb (0',1'). The step fine structure of compound {\bf 2} is
explained by exchange coupling with neighbors. It can be 
explained by the splitting of each comb into four combs (Fig. 2b).
}
\label{fig2}
\end{center}
\end{figure}

The magnetization versus applied field was recorded on single crystals
of [Mn$_4$]$_2$ using an array of micro-SQUIDs~\cite{WW_ACP_01}.
Figs. 2a and 2b show typical hysteresis loops of magnetization versus
applied field for different field sweep rates. The field is applied 
along the easy axis of magnetization of a single crystal of about
20 $\mu$m. 
These loops display step-like
features separated by plateaus. The hysteresis loops of the two 
crystals are similar. However,
compound {\bf 2} shows a fine structure that is absent in the 
hysteresis loops of compound {\bf 1}. 
We will show in the following that the main feature
of the hysteresis loops can be explained by the QTM of one Mn$_4$ molecule,
coupled by a superexchange interaction $J$ to the other unit of the 
[Mn$_4$]$_2$ dimer.
We discuss first compound {\bf 1} because the
coupling with neighboring dimers can be neglected~\cite{WW_Nature02}.
Then, we show that the fine structure observed for
compound {\bf 2} is induced by a superexchange
interaction $J'$ between neighboring dimers (Fig. 1b).

The simplest Hamiltonian
describing the spin system of an isolated SMM is:
\begin{equation}
	\mathcal{H} = -D S_{z}^2 + \mathcal{H}_{{\rm trans}}
	+ g \mu_{\rm B} \mu_0 \vec{S}\cdot\vec{H}
\label{eq_H}
\end{equation}
$S_{x}$, $S_{y}$, and $S_{z}$ are the components of the spin operator;
$D$ is the anisotropy constant defining an Ising type of anisotropy;
$\mathcal{H}_{{\rm trans}}$, containing $S_{x}$ or $S_{y}$ spin
operators, gives the
transverse anisotropy which is small compared to $D S_{z}^2$ in SMMs;
and the last term describes the Zeeman energy associated with an
effective field $\vec{H}$. For one isolated spin the effective field
is the applied field.
This Hamiltonian has an energy level spectrum with $(2S+1)$ values
which, to a first approximation, can be labelled by the quantum
numbers $M = -S, -(S-1), ..., S$ taking the $z$-axis as the quantization
axis. The energy spectrum can be obtained by using standard
diagonalization techniques. At  $\vec{H} = 0$, the levels $M = \pm S$ have
the lowest energy. When a positive field $H_z$ is applied, the levels with
$M > 0$ decrease in energy, while those with $M < 0$ increase.
Therefore, energy levels of positive and negative quantum
numbers cross at certain values of $H_z$ given by
$\mu_0 H_z \approx n D/g \mu_{\rm B}$,
where $n = 0, 1, 2, 3, ...$.
When the spin Hamiltonian contains
transverse terms ($\mathcal{H}_{\rm trans}$), the level crossings
can be {\it avoided level crossings}.
The spin $S$ is {\it in resonance}  between two states
when the local longitudinal field is close to an avoided level crossing.
The energy gap, the so-called {\it tunnel splitting}
$\Delta$, can be tuned by a transverse field
(a field applied perpendicular to the $z$ direction)
via the $S_xH_x$ and $S_yH_y$ Zeeman terms. The effect of these avoided level
crossings leads to well defined steps
in hysteresis loop measurements.

The main point to note is that the
giant spin Hamiltonian predicts always the first level crossing
at zero field, corresponding to the QTM of a SMM between $M = \pm S$
states. Thus, for compound {\bf 1} (see Fig. 2a), the single-spin
Hamiltonian is not sufficient to explain the first resonance
shifted to negative fields and the absence of the quantum
tunnelling at zero field, in contrast to other SMMs.

In order to explain the observed features in Fig. 2a, one has
to take into account the superexchange coupling $J$ between pairs of
Mn$_4$ units. A Hamiltonian for the two-coupled molecules can be written
and the energy states of the [Mn$_4$]$_2$ can be calculated by exact 
diagonalization.
More details on the dimer Hamiltonian and the corresponding Zeeman diagram
are reported elsewhere~\cite{WW_Nature02}. Here, we propose a phenomenological
model that is sufficiently simple to allow inclusion, in a second step, 
of more exchange couplings. 
The influence of the exchange coupling of the
neighboring molecule is taken into account by an exchange bias field
$H_{bias}$. The effective field $H_{z}$ acting on the molecule is
therefore the sum of the applied field $H_{app}$ and the bias field
$H_{bias}$ :
\begin{equation}
	H_{z} = H_{z}^{\rm app} + H_{z}^{\rm bias} = H_{z}^{\rm
	app} + \frac{J}{g \mu_{\rm B} \mu_0 }M_2
\label{eq_dimer}
\end{equation}
where $M_2$ is the quantum number of the neighboring molecule 
and $J$ is the associated exchange coupling. 
In the following we explain the hysteresis loops when the field 
$H_{z}^{\rm app}$ is swept from
negative to positive values.
At low temperature,
$M_2$ has two possible values $M_2 = \pm S = \pm 9/2$.
We expect therefore
resonant QTM for applied fields $\mu_0 H_{z}^{\rm app}
\approx n D/g \mu_{\rm B} \pm M_2 J/g \mu_{\rm B}$,
where $n = 0, 1, 2, 3, ...$. The two possibilities of $M_2$
are represented by two combs in Fig. 2a. The first comb (0,1,2)
corresponds to $M_2 = -9/2$ and the second one (0',1') to $M_2 = 9/2$.
This model describes all observed quantum transitions in Fig. 2a
with two fitting parameters $D/k_{\rm B} = -0.72$ K and 
$J/k_{\rm B} = 0.1$ K. It neglects co-tunneling 
and other two-body tunnel transitions having a lower probability of 
occurrence~\cite{WW_Nature02,WW_PRL02}.

\begin{figure}
\begin{center}
\includegraphics[width=.42\textwidth]{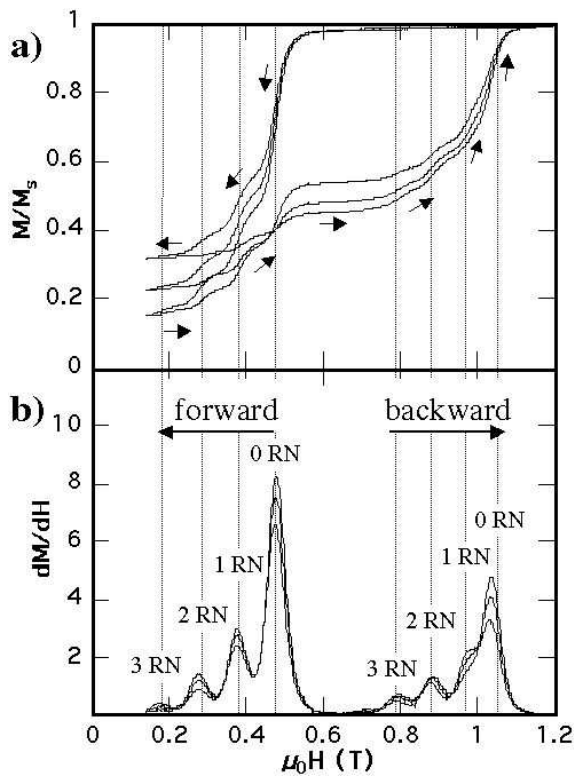}
\caption{Field sweep rate dependence of (a) the minor hysteresis loops and
 (b) the derivatives of the hysteresis loops, measured 
 on a single crystal of compound {\bf 2} at 0.04 K.
 The positions corresponding to 0, 1, 2, or 3 reversed neighbors (RN) 
 are indicated.}
\label{fig3}
\end{center}
\end{figure}

Compound {\bf 2} displays hysteresis loops (Fig. 2b) similar to 
those of compound {\bf 1}. However, the total exchange coupling is larger
for compound {\bf 2}. The values of $D/k_{\rm B} =-0.75$ K 
and $J/k_{\rm B}$ = 0.15 K
were obtained from the field positions of the steps in the hysteresis loops.
Another difference between the two compounds is that the hysteresis loops of
compound {\bf 2} exhibit fine structure that can not be explained
by the dimer model described above (Eq. 2).
In order to better analyze this fine structure, minor hysteresis
loops were measured (Figs. 3 and 4). First the sample is
saturated in positive field; all the molecules are in the $M = +9/2$ state.
Then the field is decreased. The system approaches
the first avoided energy level crossing at a field value
of $\approx$0.5 T. 
A fraction of the dimers switches from +9/2 
to $-$9/2, and the total 
magnetization of the system
decreases, generating a step in the hysteresis loop. When the
magnetization reaches the second 
plateau ($\approx$0.2 T), the field is swept back
towards positive saturation; the tunneling 
from $M = -9/2$ to 9/2 is favored via the
exited state 7/2 ($\approx$1 T).
After this transition the sample
reaches positive saturation. 
The purpose of these minor hysteresis loops
is to confirm the fine structure of each
transition starting from different initial states.

\begin{figure}
\begin{center}
\includegraphics[width=.42\textwidth]{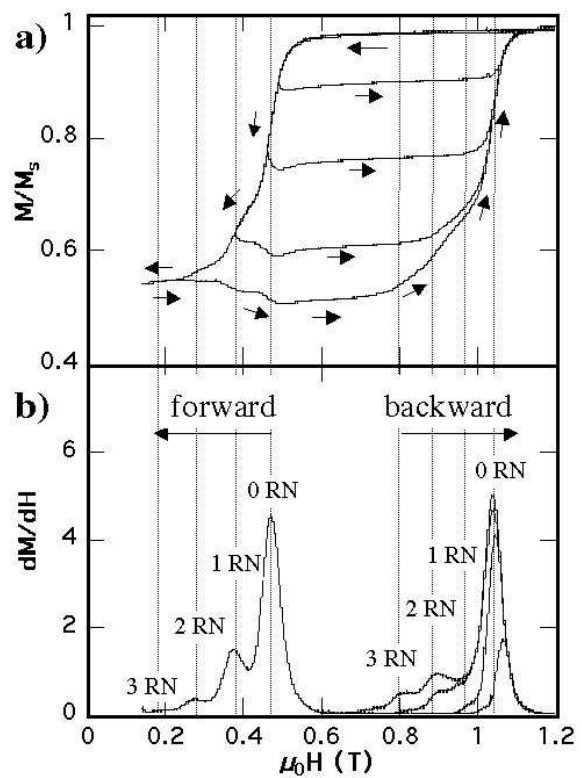}
\caption{(a) Several minor hysteresis loops and
 (b) their derivatives, measured 
 on a single crystal of compound {\bf 2} at 0.04 K. 
 The field sweep rate 
 is 0.14 T/s. The positions corresonding 
 to 0, 1, 2, or 3 reversed neighbors (RN) 
 are indicated.}
\label{fig4}
\end{center}
\end{figure}

The tunnel transitions exhibit four 
equidistant kinks, that we explain by the
exchange coupling to the three neighboring dimers. 
The spin of the three neighboring molecules 
can be either aligned with the magnetic field or reversed, 
leading to four different situations: from zero to three
reversed neighbors.

The exchange coupling between a molecule and its neighbors acts
like a supplementary field bias and shifts further
the resonance fields. The total field bias 
induced by the neighbors and the
other Mn$_4$ unit of the dimer can be written:
\begin{equation}
	H_{\rm bias}^{tot} = \frac {1}{g\mu_{B}\mu_{0}}
	\left( JM_{2}+\sum_{i=1}^{3}J'M'_{i}\right)
\label{eq_dimer_2}
\end{equation}
where the first term is the contribution of the intradimer coupling, 
and $M'_{i}$ is the quantum number of the
three neighboring dimer molecules $i$ (Fig. 1b).

After positive saturation all the molecules
are aligned with the field. The first kink in the hysteresis loop 
corresponds to the QTM of one molecule 
in the bias field of its non-reversed neighbors.
The resonance is shifted towards negative values by the bias field
$H_{\rm bias} = 9/(2 g\mu_{B}\mu_{0}) (J+3J')$ (see Eq.4). 
After this first kink,
some molecules now have one reversed neighbor. At the
second kink it is this newly created population which tunnels generating
molecules with two reversed neighbors. The corresponding field shift 
is $H_{\rm bias} = 9/(2 g\mu_{B}\mu_{0}) (J+J')$. The third and
the fourth kinks are generated by the QTM of molecules having,
respectively, two and three reversed neighbors. The field shift between two 
consecutive kinks is $\approx$0.1 T,
corresponding to an interdimer interaction $J' \approx$ 0.015 K.

Minor hysteresis loops were measured for different field sweep 
rates (Fig. 3) and reversal fields (Fig. 4) in order to probe the 
step heights of the fine structure: 
the smaller the sweep rate the higher the
resulting kink. This dependence is justified by the Landau Zener model.
The main point to note is that heights of two consecutive kinks are 
correlated. The second
kink height is smaller than the first kink
height, the third smaller than the second, and so on. 
This result is in good agreement with
our model: in order to have quantum tunneling of molecules with $n$ reversed
neighbors, the $n$ neighbors must have previously reversed.

All the other transitions exhibit the same kind of fine
structure, which can be explained by the above model leading to
the 8 combs in Fig. 2b, giving for the three
fitting parameters $D/k_{\rm B} \approx -0.75$ K, 
$J/k_{\rm B} \approx 0.1$ K
and $J'/k_{\rm B} \approx 0.015$ K.

The above results demonstrate that a three dimensional network of
exchange coupled SMMs doesn't suppress QTM. The intermolecular 
interactions are strong enough to cause a clear field bias,
but too weak to transform the
spin network into a {\it classical} antiferromagnetic
material. This three dimensional network
of exchange coupled SMMs demonstrate that the QTM can be controlled using
exchange interactions, and opens up new
perspectives in the use of supramolecular chemistry
to modulate the quantum physics of these
molecular nanomagnets.

%\bibliographystyle{wernsdor}
%\bibliography{wernsdor}

\end{document}